# Photocatalytic Activity of Pulsed Laser Deposited TiO$_2$ Thin Films

H. Lin[1], Abdul K. Rumaiz[2], Meghan Schulz[3], Demin Wang[1], Reza Rock[4], C.P. Huang[1] and S. Ismat Shah[2,3],

1. Department of Civil and Environmental Engineering, University of Delaware, Newark, Delaware, 19716, USA.
2. Department of Astronomy and Physics, University of Delaware, Newark, Delaware, 19716, USA.
3. Department of Material Science Engineering, University of Delaware, Newark, Delaware, 19716, USA.
4. Department of Chemical Engineering, University of Delaware, Newark, Delaware, 19716, USA.

Nanostructured TiO$_2$ thin films were prepared by pulsed laser deposition (PLD) on indium doped tin oxide (ITO) substrates. Results from X-ray photoelectron spectroscopy (XPS) show that Ti 2p core level peaks shift toward the lower binding energy with decrease in the buffer gas pressure (O$_2$:Ar = 1:1). This suggests that oxygen vacancies are created under insufficient oxygen conditions. Anatase to rutile ratio is also found to be system pressure dependent. Under deposition pressure of 750 mTorr only anatase phase was observed even at 1073 K substrate temperature which is much higher that the bulk anatase to rutile phase transformation temperature. The deposited TiO$_2$ thin films were fabricated as photoanodes for photoelectrochemical (PEC) studies. PEC measurements on TiO$_2$ photoanodes show that the flatband potential (V$_{fb}$) increases by 0.088 eV on absolute vacuum energy scale (AVS) with decrease in the deposition pressure, from 750 to 250 mTorr at 873K. The highest incident photon to current conversion efficiency [IPCE($\lambda$)] of 2.5 to 6 % was obtained from the thin films prepared at substrate temperature of 873K. Combining the results from XPS and PEC studies, we conclude that the deposition pressure affects the concentration of the oxygen vacancies which changes the electronic structure of the TiO$_2$. With reference to photoelectrochemical catalytic performance, our results suggest that it is possible to adjust the Fermi energy level and structure of TiO$_2$ thin films by controlling the buffer gas pressure and temperature to align the energy of the flatband potential ($V_{fb}$) with respect to specific redox species in the electrolyte.

# 1. Introduction

Titanium dioxide ($TiO_2$) has been proven to be an effective material for applications such as photocatalysis [1, 2], dye sensitized solar cells [3, 4], heterogeneous catalysis [2, 5], self-cleaning/antifogging surface coatings [6], etc. Typically, $TiO_2$ is used in two main forms: powder and thin film. The powder form of crystalline $TiO_2$ is commonly used for gas and liquid phase catalysis. Its photocatalytic activity is normally determined by the particle size [2], phase composition [7], and the position of the conduction and valance bands in the energy scale [8]. In the thin film form, $TiO_2$ is usually used in photovoltaic applications such as photoelectrochemical system (PEC) and dye sensitized solar cell (DSSC) for photon harvesting [4, 9]. Moreover, $TiO_2$ in the thin film form offers the advantage of energy alignment between the energy position of the valance band edge and the redox species in the electrolyte by potential biasing the photoanodes. This helps to optimize the quantum efficiency.

Several $TiO_2$ thin film deposition techniques have been reported which include metalorganic chemical vapor deposition (MOCVD) [10, 11], sol-gel [12], electrophoretic deposition [13], reactive rf sputtering [14], and pulsed laser deposition (PLD) [15-21]. Among available techniques, PLD is a high energy process which provides a well adherent thin film with good mechanical rigidity [22] and surfaces with high specific surface area [19]. In addition, PLD also offers advantages such as stoichiometrically transferring material from target to the substrate [22], capability of inert and reactive gas deposition, wide range of operational pressure and temperature, and variety in options for substrate materials.

TiO$_2$ thin film prepared by pulsed laser deposition has been studied by various research groups [9, 15-17, 20, 21]. However, different type of targets (i.e. Ti and TiO$_2$), variation in substrate materials, wide range of operation pressures (i.e. from ultra high vacuum to 600 mTorr), and the differences in synthesis temperature (i.e. from room temperature to 1273K) make it difficult to compare and understand the differences in properties of the thin films in a consistent manner. The anatase and rutile multi phase structures in TiO$_2$ thin films were observed by several groups [9, 15, 21]. Only Luca et. al. [21] have addressed the increase in anatase phase composition with increase in oxygen partial pressure. However, the possible mechanism was not proposed. The distinctive difference of PLD process from chemical based synthesis (i.e. Sol-Gel and MOCVD) is that multi-valance Ti species are usually created during the PLD deposition process. This is unlikely to occur in chemical synthesis methods due to the presence of oxygen molecular in precursor itself and it is normally operated at oxygen rich environment [23]. Even with TiO$_2$ target, Kitazawa [24] observed the presence of neutral and ionized Ti and TiO species, based on the optical measurement of the laser ablation plume. The reduced TiO$_2$ surface (oxygen vacancy) is of particular interest for its photocatalysis applications such as dehydration of formic acid and dissociation of water molecular [25-27]. Very few studies have discussed the relationship between the valance states of Ti and the deposition condition during PLD synthesis of TiO$_x$ thin films [21]. Luca et. al. [21] reported that O:Ti ratio ranges from 1.78 to 2.0 at different combination of temperatures and pressures. They also reported that the TiO$_2$ and its suboxides (i.e. TiO and Ti$_2$O$_3$) co-exist when deposition was performed at 423K.

Pulsed laser deposition of TiO$_2$ is a simple process yet involves a complicated physical phenomenon. Variation in deposition parameters such as pressure and temperature results in different chemical and structural compositions [15, 18, 21]. Among the reports on pure TiO$_2$ thin films prepared by PLD methods, only few have tested the photocatalytic activities of the films [19, 28].

In this study we prepared TiO$_2$ thin films by PLD method at substrate temperatures of 873 and 1073K to avoid the formation of amorphous TiO$_2$ [18]. Buffer gas (O$_2$:Ar = 1:1) pressure, from 250 to 750 mTorr, was used to prevent the formation of suboxide TiO or Ti$_2$O$_3$ [21]. By operating the PLD system within this temperature and pressure regimes, we were able to prepare slightly reduced to stoichiometic TiO$_2$ films with a resulting variation in the rutile/anatase phase composition. For photovoltaic applications, conductive substrate is required. Quartz substrate coated with indium doped tin oxide (ITO), which is one of the most widely used transparent conductive oxide (TCO), were used as the substrates in our study. The surface topography, crystalline structure, binding energy of Ti, quantum efficiency spectra, and photocurrent onset of TiO$_2$/ITO thin films prepared under different synthesis conditions of temperature and pressure are reported in this paper.

## 2. Experimental Procedure

*2.1 Film Preparation*

The schematic diagram of the experimental setup is shown in figure 1. A KrF excimer pulsed laser system (wavelength λ =248 nm) was used for the deposition of $TiO_2$ thin film. A turbo molecular pump and a mechanical pump were employed in series to maintain system base pressure at $1.0 \times 10^{-6}$ Torr. Buffer gas of 50:50 by volume $Ar:O_2$ mixture was used for the reactive PLD process. Two 500W halogen lamps were used as irradiative heating source to control the substrate temperature from room temperature up to 1073K. The target was prepared by sintering pure $TiO_2$ powder (Aldrich) at 1273K $TiO_2$ disk was used as the rotating target, which only consists of rutile structured $TiO_2$. Target rotation speed was kept at 15 rpm. The incident laser beam maintained a 45º angle to the target surface. Laser pulse frequency was set at 15 Hz with a calculated laser beam fluence of 1.8 $J/cm^2$. The deposition rate of $TiO_2$ thin films was about 0.09 Å per laser pulse. Indium doped Tin oxide (ITO) coated quartz with the sheet resistance of 10±1 Ω/□ (SPI supplies Inc., PA, USA) was used as the substrate for deposition. Prior to the deposition, all ITO coated quartz substrates were cleaned ultrasonically in pure acetone solution and triple rinsed with deionized water (18.0 MΩ). Three series of $TiO_2$ thin film were deposited corresponding to substrate temperature of (1) room temperature, (2) 873K, and (3)1073K. 873K and 1073K correspond to below and above the reported anatase to rutile phase transformation temperature (~973K) for bulk $TiO_2$ [29, 30]. Depositions were carried out at buffer gas pressures of 250 and 750 mTorr for each substrate temperature.

*2.2. Film Characterization*

The surface topography of TiO$_2$ thin films was observed by using atomic force microscopy (AFM) which was equipped with a piezoelectric tube scanner. Images were taken under contact mode with a scan rate of 1.11 Hz (J-scanner, multimode AFM/SPM, Veeco). The crystalline structure of TiO$_2$ thin films was determined by using Rigaku D-Max B diffractometer which was equipped with a graphite crystal monochromator. θ−2θ scans were recorded using Cu $K_\alpha$ radiation of wavelength of 1.5405 Å from 20º to 80º with a step size of 0.05º. The oxidation state of Ti was determined by XPS. An SSI-M probe XPS was used employing Al $K_\alpha$ (hν=1486.6 eV) excitation source. High resolution XPS spectra were collected at 26 eV pass energy with a dwell time of 100 ms per point. Peak positions were referenced to the C 1s peak at 284.6 eV. The binding energies of the Ti 2p$_{1/2}$ and 2p$_{3/2}$ were examined.

*2.3. Photocurrent Measurement*

Deposited TiO$_2$/ITO thin films were fabricated as the photoanode for the photoelectrochemical (PEC) studies. The photoelectrochemical system was equipped with a three-electrode potentiostat (model AFRDE 4, Pine Instrument Inc., USA), a custom made photoelectrochemical cell with fused silica window, and a monochromatic excitation source (Model RF-5301, Shimadzu, Japan). Platinum wire was used as the counter electrode. Saturated calomel electrode (SCE) was selected as the reference electrode.

Sweeping voltage was within the range of ± 0.6 Volts (vs. SCE). Potassium iodide [KI = 0.05 M, (pH~9)] was used as electrolyte our photocurrent measurement. Photocurrent was recorded within the irradiation wavelength ($\lambda$) range between 300 and 500 nm for the incident photon to current conversion efficiency [IPCE($\lambda$)] computation.

## 3. Results and Discussion

Figure 2 shows the AFM images of the $TiO_2$ thin films that were deposited at substrate temperatures of 873 and 1073K and deposition pressures of 250 and 750 mTorr, respectively. From the topographic images it can be seen that for the same pressure, the topography of the films deposited at 873K appears to be more uniform than the topography of the sample deposited at 1073K. The section analysis shows that RMS roughnesses are 8.4 and 11.2 nm for thin films deposited under 250 and 750 mTorr, system pressure respectively, at T = 873K. At T = 1073K, the RMS roughnesses are much higher, 47.2 and 50.6 nm, for deposition under 250 and 750 mTorr, respectively. There is a very small pressure dependence of the roughness but temperature certainly changes the topography drastically. A possible explanation for this observation is that surface mobility of the adatoms is higher at higher temperature (1073K) which results in higher surface diffusion length, island separation, and lateral size. When island separation length is greater than lateral size of the island, terrace and stairs topography are normally favored. On the other hand, surface mobility of the adatoms is lower at lower temperature (873K), islands are more closely spaced. If island separation is smaller than the island lateral size, more uniformed growth of the thin film is preferred [31, 32].

The Ti core level electron binding energy for the $TiO_2$ thin films deposited at different temperature (T = 873 and 1073K) and buffer gas pressure (P = 250 and 750 mTorr) are compared in Figure 3. At 873K, the binding energy of Ti $2p_{3/2}$ increases from 457.9 to 458.2 eV when system pressure increased from 250 to 750 mTorr (Figure 3-a). The binding energy of Ti $2p_{3/2}$ in stoichiometric $TiO_2$ has been reported to be in the range of 458.5 to 458.9 eV [33, 34] and the binding energy of Ti $2p_{3/2}$ in $Ti_2O_3$ has been reported to be around 456.8 to 456.9 eV [34, 35]. Evidently, the binding energies of Ti $2p_{3/2}$ in these two samples are much closer to that reported value for $TiO_2$ than for $Ti_2O_3$. The 0.3 eV increase in binding energy from 457.9 to 458.2 and relatively larger peak separation between Ti $2p_{1/2}$ and Ti $2p_{3/2}$ for films deposited at 250 mTorr suggest that higher Ti valance states was created at higher oxygen partial pressure. The same trend was observed for samples prepared at 1073K (Figure 3-b). The binding energy of Ti $2p_{3/2}$ increases from 458.1 to 458.5 eV with the increase in system pressure from 250 to 750 mTorr. Note that the binding energy of Ti $2p_{3/2}$ at 458.5 eV is within the reported binding energy range for stoichiometric $TiO_2$. The buffer gas pressure can contribute to the differences in propagation characteristics of ablated species ($TiO_x$ atoms, ions and clusters) [17]. Our results suggest that at the same temperature, higher concentration of oxygen vacancy is generated at lower pressure condition which is likely due to insufficient availability of O species at lower pressure conditions.

The Figure 4-a and b show the high resolution XRD scan (2θ = 22° to 30°) of the prepared TiO$_2$ thin films deposited at temperature of 873 and 1073K, respectively. As can be seen, the anatase and rutile phases co-exist through the pressures between 250 to 500 mTorr at both temperatures. When deposition pressure increased to 750 mTorr, there was no clear rutile R(110) peak from our XRD results for samples deposited at either temperatures. The anatase weight fraction (figure 4-c) is computed, based on Spurr's formula (equation 1) [21, 36] where the I$_A$ and I$_R$ are the integrated intensity of anatase A(101) and rutile R(110) peaks from the XRD pattern, respectively.

$$W_A = \frac{I_A}{I_A + 1.265 I_R} \quad \ldots\ldots\ldots\ldots\ldots\ldots\ldots\ldots (1)$$

Our results show that the anatase contents increase with increase in the buffer gas pressure. A possible explanation could be that the anatase has a higher c/a ratio than rutile. The c/a ratio for anatase and rutile phases are commonly reported with value of 2.51 and 0.64, respectively [37]. Under lower pressure conditions (i.e. 250 mTorr), insufficient oxygen causes vacancy formation and destabilizes the larger c/a anatase structure. Even if anatase structure survives at low oxygen pressure conditions, certain concentration of vacancies is expected. In our previous work at very low partial pressure of O, we have seen a shoulder in the Ti 2p peaks related to the formation of vacancies [38]. In the current case the concentration of vacancies is not high enough to register an effect on the XPS spectra. At higher buffer gas pressure (i.e. 750 mTorr), sufficient oxygen is supplied compared to the former case. Under such conditions the required c/a of anatase can be stabilized and a more stoichiometric material is obtained due to a lower vacancy concentration. Similar trend was also reported by Luca [21] and Kitazawa [18] except under different temperature, pressure, and substrate conditions. Our results show that even when the thin film was prepared at temperatures higher than what is generally considered to be the anatase → rutile phase transformation temperature, 973 °K, the anatase phase is still stable.

The flatband potential of a semiconductor photoelectrode can be determined based on photocurrent onset potential ($E_{onset}$) [39]. Based on the linear scan voltammetry, the Tafel plots obtained from $TiO_2$ thin films deposited at different temperature and buffer gas pressures are shown in figure 5. The photocurrent onset corresponds to the peak on the left hand side of each curve. To reference the applied potential, we plot the voltage scale in terms of the normal hydrogen electrode potential (NHE) and the absolute vacuum scale (AVS) with respect to log|J| (J: current flux under irradiation, $mA/cm^2$) . It can be seen that at 873K, the flatband potential ($V_{fb}$) of $TiO_2$ electrode are -0.205 and -0.117 V (v.s. NHE) for thin films prepared under 250 and 750 mTorr buffer gas pressures, respectively. That is, the flatband potential ($V_{fb}$) is increased by 0.088 eV in absolute vacuum scale when deposition pressure decreased from 750 to 250 mTorr. The flatband potential is related to the potential and charge distribution at the electrode-electrolyte interface which is affected mainly by the surface states of semiconductor and the adsorbed ions at the surfaces [40]. We presume that the main factor affecting $V_{fb}$ is the variation in electronic properties of the $TiO_2$ thin films since all experiments were performed under identical conditions. We have earlier described the formation of reduced $TiO_2$ under certain experimental condition due to the creation of oxygen vacancies during deposition. The valance band mainly has O derived 2p states separated from the empty Ti derived 3d states by a bulk bandgap of 3.2 eV. [41, 42]. By removing one $O^{2-}$ ion from $TiO_2$ lattice, two electrons are freed and they will potentially occupy the Ti 3d orbitals to form a more localized states within the band gap [43]. This has been predicted in several DFT calculations [42, 44] and also observed experimentally [38]. Thus when the Ti derived conduction band becomes occupied upon reduction, it has been shown that the O 2p valence band moves away from the Fermi level [38]. The easiest way to interpret this would be to assume a rigid band model where the reduction moves the Fermi level into the conduction band of $TiO_2$. Therefore, it is reasonable to see higher $V_{fb}$ for more reduced $TiO_2$ due to reasons mentioned earlier. For samples prepared at 1073K, the $V_{fb}$ is positioned at +0.022 and +0.011 V (v.s. NHE) for thin films prepared under 250 and 750 mTorr buffer gas pressure, respectively. Although the variation of $V_{fb}$ at this temperature has reverse trend compared to 873K prepared samples, the 0.011 eV difference is too small to surpass the experimental error. At both substrate temperature, XPS results show 0.3 ~ 0.4 eV variation in binding energy of Ti $2p_{3/2}$ core level when varying the deposition pressure between 250 and 750 mTorr. However, the difference in flatband potential ($V_{fb}$) is only pronounced for thin films prepared at the substrate temperature of 873K. Since the XPS is a surface sensitive technique, we suspect that the binding energy difference between samples prepared under 250 and 750 mTorr at 873K is greater in the bulk than at the surfaces. In other word, our XPS results might have under-estimated the binding energy difference due to possible oxidation of the reduced $TiO_2$ surface during the sample transfer process.

To examine the photocatalytic efficiency of prepared TiO$_2$ thin films, the incident photon conversion efficiency [IPCE($\lambda$)] test was performed. The [IPCE($\lambda$)] is defined as the number of electron transferred per incident photon (equation 2) [14]

$$IPCE(\%) = \frac{J_{sc}(\lambda)}{eI_{inc}(\lambda)} = \{[1240 \times J_{sc}(A/cm^2)]/[e \times \lambda(nm) \times I_{inc}(W/cm^2)]\} \times 100 \dots\dots\dots\dots(2)$$

where the $J_{sc}(\lambda)$ is short-circuit photocurrent density under monochromatic light (A/cm$^2$), $e$ is the electron charge (Coulomb), $I_{inc}(\lambda)$ is the incident photon flux (W/cm$^2$).

This method provides more information than optical absorption measurement does due to factors such as the surface morphology and mass transfer limitation effect (in the presence of electrolyte) which are not fully taken into account in the later method. The incident photon to current conversion efficiency [IPCE($\lambda$)] of TiO$_2$ thin films are compared in figure 6 in the spectrum range from $\lambda$ = 300 and 500 nm. The highest [IPCE ($\lambda$)] obtained is in the range of 5~5.5 and 0.4~1.8 % for 873 and 1073K prepared sample, respectively (at $\lambda$ = 320 nm). Evidently, the [IPCE($\lambda$)] is much higher at deposition temperature of 873K than that at 1073K. A possible reason is that TiO$_2$ thin films deposited at 1073K is likely to be more densely deposited compare to that prepared at 873K. Thus, the films prepared at lower temperature prepared sample (T = 873K) may have relatively more porous structure which allows more reactive surfaces for photocatalytic reaction to take place. In figure-6, it is shown that the [IPCE($\lambda$)] spectra for films deposited at 250 mTorr (consist mainly Rutile) extends into visible range (>400) but gives relatively lower photon conversion efficiency in deep UV range (< 350 nm) at both temperature. For films prepared at 750 mTorr (consist mainly anatase), the [IPCE($\lambda$)] spectra gives greater photon conversion efficiency than that at 250 mTorr in deep UV range. However, the photon conversion efficiency becomes noticeable only in UV range ($\lambda$ < 400 nm). Our [IPCE(l)] spectra is very consistent with the reported band gap values. Note that the band gaps of the anatase and rutile have been reported to be 3.2 ($\lambda$ = 390 nm) and 3.0 ($\lambda$ = 410 nm) eV, respectively [45]. It also has been proven experimentally that anatase has significantly higher light absorption capacity than that of rutile [45]. The difference in [IPCE($\lambda$)] spectra is clearly due to the phase composition of the $TiO_2$ thin films.

## 4. Conclusion

Nano-structured $TiO_2$ thin films were deposited on ITO substrates by pulsed laser deposition (PLD) under different conditions of temperatures and pressure (T = 873 and 1073K; P = 250 and 750 mTorr). AFM results show that samples prepared at 873K have much more uniform surfaces and smaller particle size than that prepared at 1073K. Our XPS results indicate the binding energy of Ti 2p core level are system pressure dependent, which suggest that oxygen vacancies can be created depending on experimental condition. At 250 and 500 mTorr, it was found that both anatase and rutile phases co-exist regardless of the deposition temperature. However, under 750 mTorr, only anatase phase was observed even at the temperature higher than commonly reported anatase → rutile phase transition range (700 °C). Based on photoelectrochemical studies, we found that the flatband potential of $TiO_2$ thin films is increased in absolute vacuum energy scale with decrease in operational pressure at deposition temperature of 873K. Combining the results from XPS, we conclude that the shift in $V_{fb}$ is induced by oxygen vacancy which results in changing in the electronic structure of reduced $TiO_2$. The incident photon to current conversion efficiency [$IPCE(\lambda)$] was observed to be much higher for thin films deposited at T = 873K than T = 1073K. This is likely due to more porous surface is created at lower deposition temperature condition. As a strategy towards photoelectrochemical catalytic performance, our results suggest that one can manipulate the Fermi level energy and structure of $TiO_2$ thin films by controlling the buffer gas pressure and temperature to align the energy of the flatband potential ($V_{fb}$) with respect to specific redox species in electrolyte.

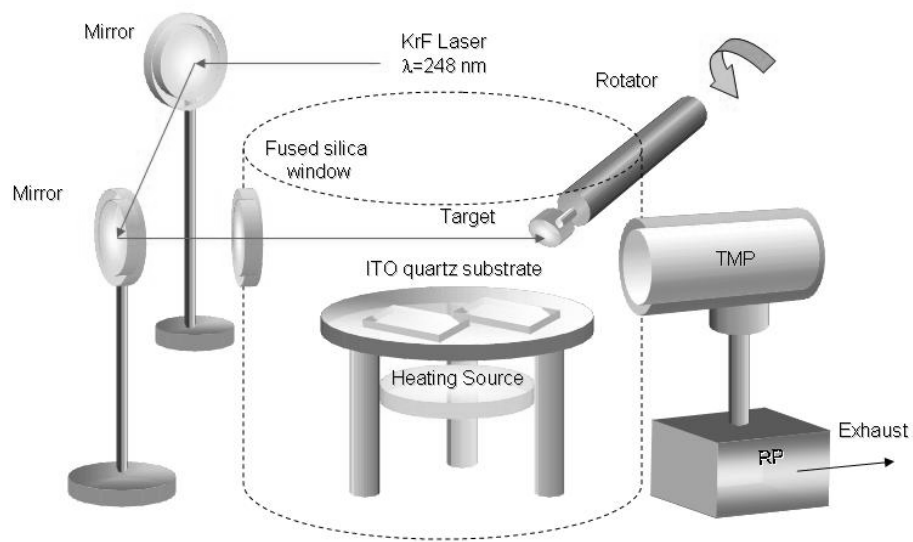

**Figure 1. Schematic diagram of the pulsed KrF gas excimer laser PLD system.**

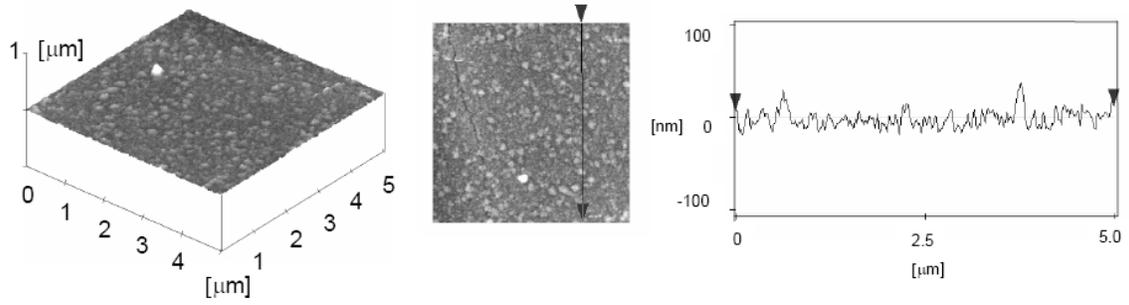

(2-a) 873K, 250 mTorr

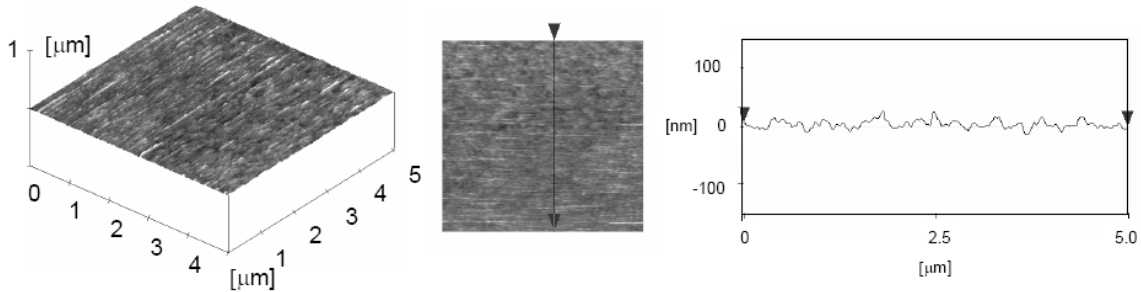

(2-b) 873K, 750 mTorr

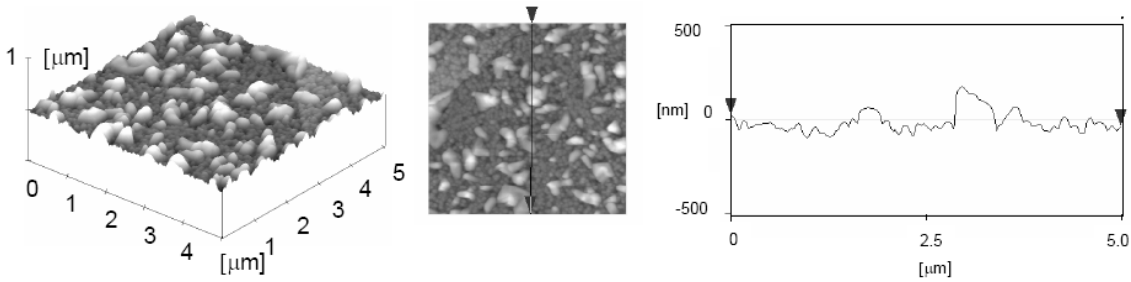

(2-c) 1073K, 250 mTorr

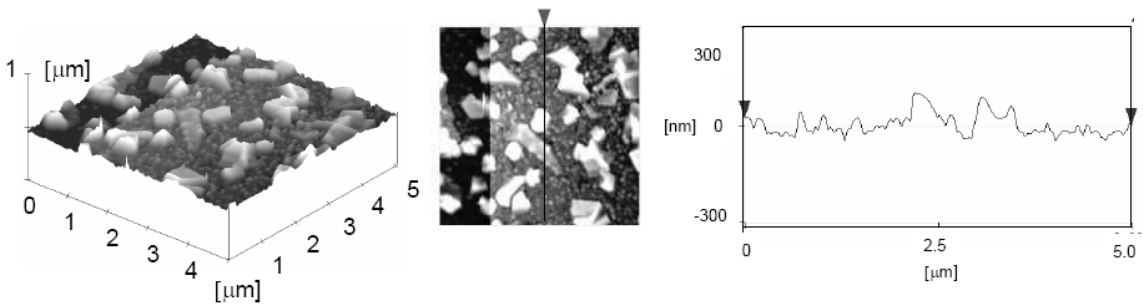

(2-d) 1073K, 750 mTorr

**Figure 2.**  AFM images of the TiO$_2$ thin films deposited at temperature of 873 and 1073K under system pressure of 250 and 750 mTorr.

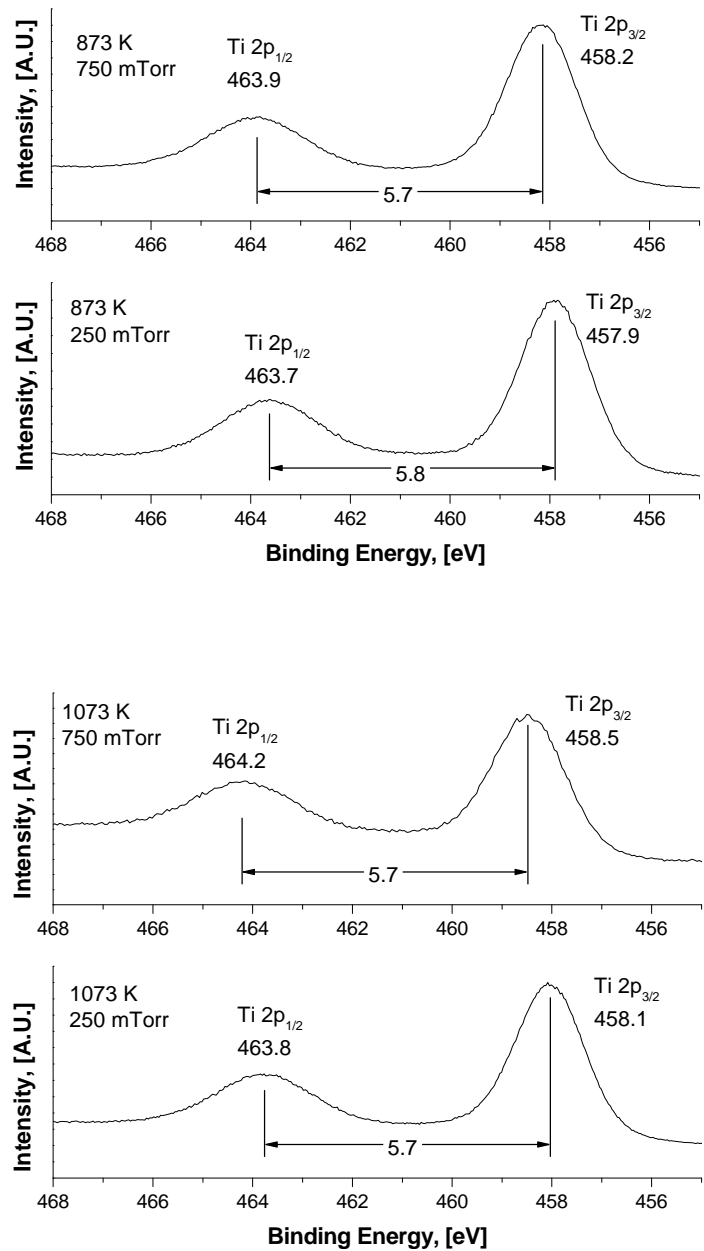

**Figure 3.    High Resolution XPS spectra of Ti 2p region.**

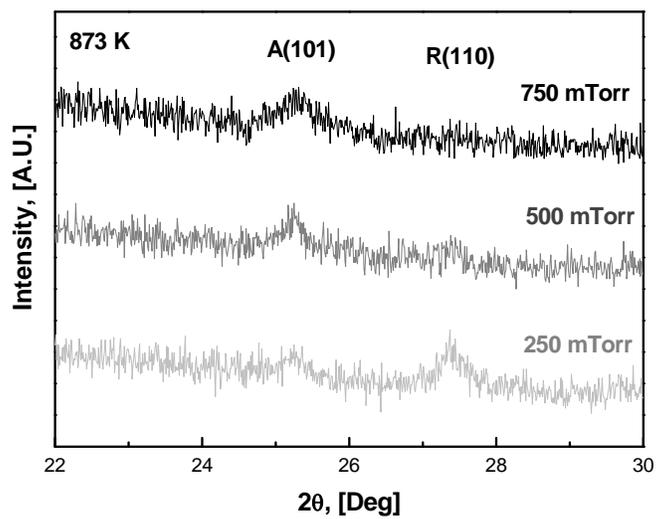

(4-a)

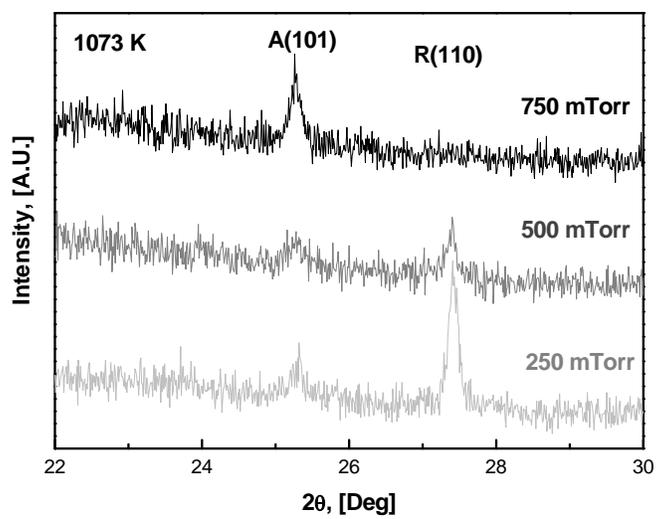

(4-b)

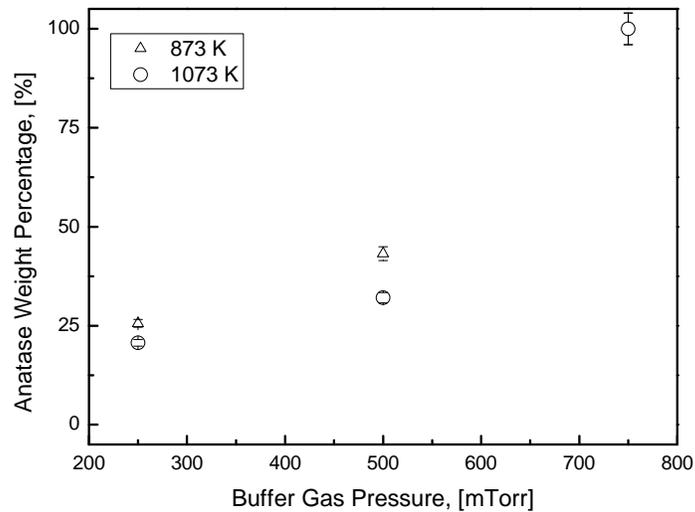

(4-c)

**Figure 4.  XRD results of as-deposited TiO$_2$ thin films. (a) High resolution scan of A(101) and R(011) region at substrate temperature of 873K (b) High resolution scan of A(101) and R(110) region at substrate temperature of 1073K, and (c) Anatase weight percentage and its corresponding operation pressure at different temperature.**

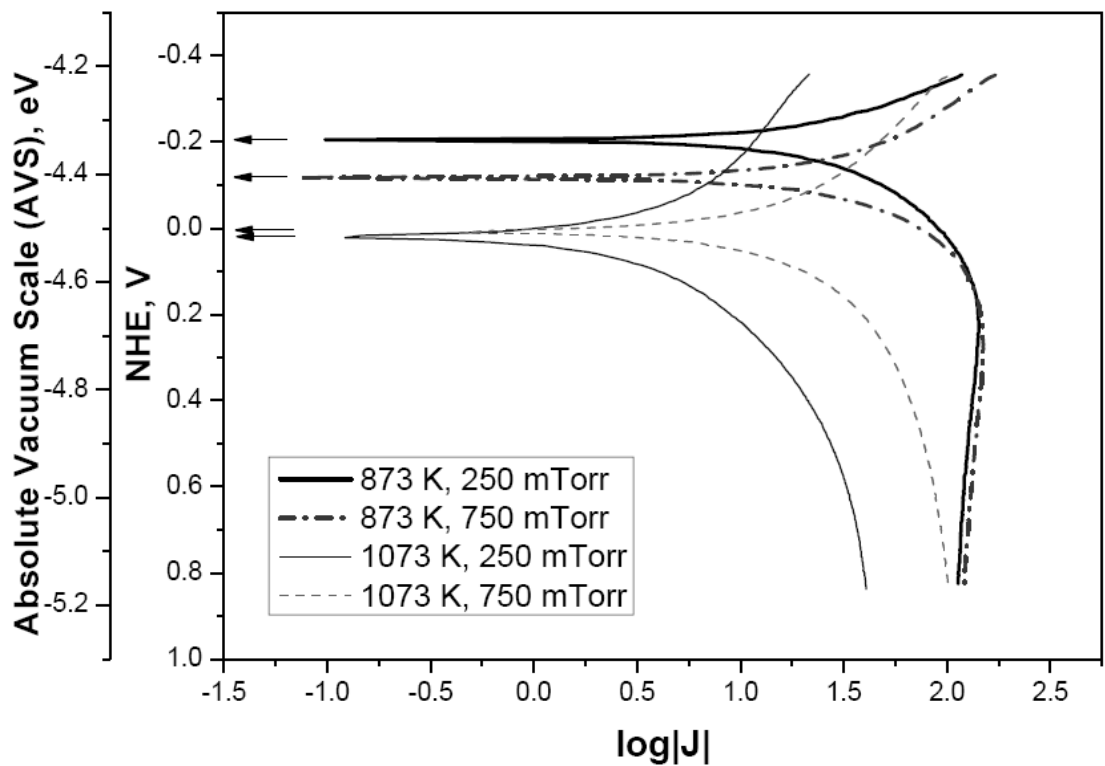

**Figure 5**      **Tafel plots of TiO$_2$ thin films prepared under different buffer gas pressure and temperature. The applied voltage is converted with reference to the standard normal hydrogen electrode (NHE) and the absolute vacuum scale (AVS).**

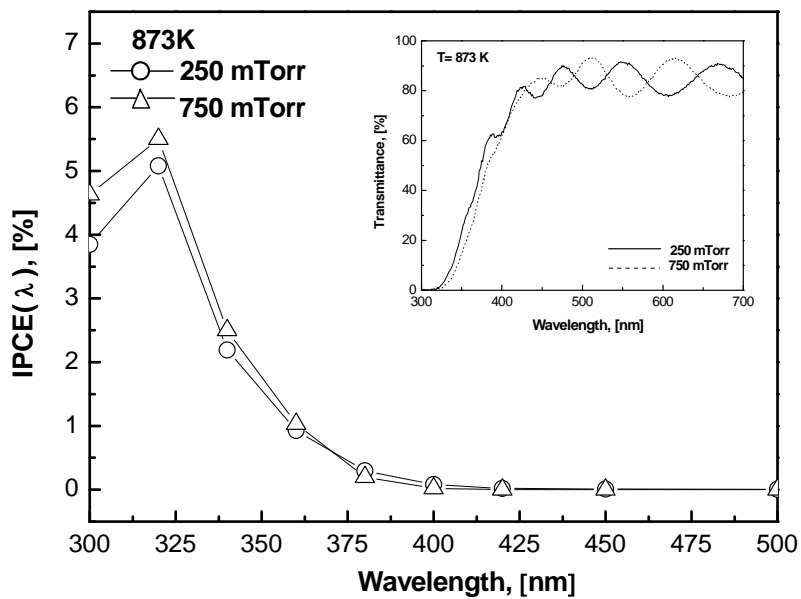

(6-a)

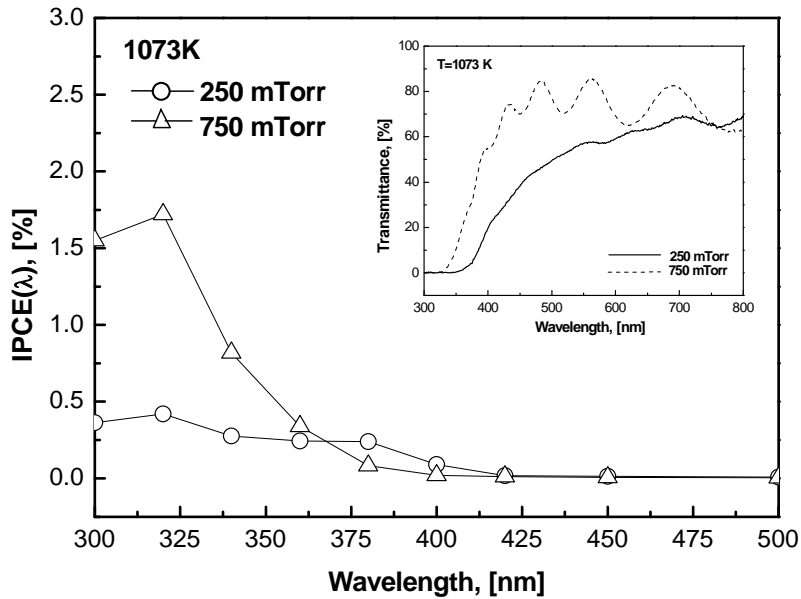

(6-b)

**Figure 6** Incident photon to current conversion efficiency [IPCE($\lambda$)] between wavelength of 300 to 500 nm. Inserts are the corresponding transmittance measurements obtained by double beam UV-VIS spectroscopy.